\documentclass[11pt]{article}
\usepackage{blois2002,epsfig}
\def\be{\begin{equation}}
\def\ee{\end{equation}}
\def\bes{\begin{eqnarray*}}
\def\ees{\end{eqnarray*}}
\def\bea{\begin{eqnarray}}
\def\eea{\end{eqnarray}}

\def\ktpp{\ensuremath{K \to \pi \pi \,}}
\def\ktp{\ensuremath{K \to \pi \,}}
\def\cpt{\mbox{CHPT\,}}
\def\no{\nonumber \\}
\def\co{\; \; ,}
\def\fs{\; \; .}

\begin{document}
\vspace*{4cm}
\title{DISPERSIVE TREATMENT OF \ktpp}

\author{ M.B\"UCHLER}

\address{Institute for Theoretical Physics, Winterthurerstr. 190,\\
Zurich 8057, Switzerland}

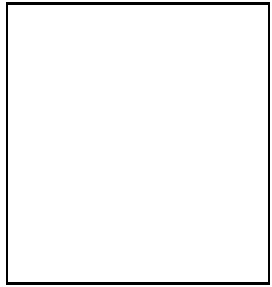
\maketitle
\abstracts{
We discuss a new method to treat the \ktpp amplitude dispersively, 
taking into full account the effects of final state interatcions. 
Our approach is based on a set of dispersion 
relations for the \ktpp amplitude, in which the weak Hamiltonian 
carries momentum. In these dispersion relations two subtraction constants 
have to be introduced, whereby one can be related via a soft pion theorem to 
the \ktp amplitude. The second is presently unknown, and we use 
lowest order Chiral Perturbation Theory for a first guess of it's value. 
We emphasize the advantage of combining  this approach 
with lattice input which could provide the two subtraction constants 
with sufficient accuracy.}

\section{Introduction}
Lattice QCD would in  principle be the appropriate tool for 
the calculation of the \ktpp ampli\-tude, accounting fully for its 
nonperturbative nature. 
In practice, an explicit 
calculation of \ktpp is not yet feasible, although there was 
some progress in this direction recently \cite{Lellouch:2000pv}. 
The major obstacle for direct lattice calculations were already 
identified some time ago by Maiani and Testa in a no-go  
theorem, stating that for decays into two or more particles, their 
interaction with each other makes a direct calculation on the 
lattice impossible \cite{Maiani:ca}. The standard workaround for \ktpp is 
the calculation of the unphysical \ktp matrix element, and the use of 
a relation valid at lowest order in chiral perturbation theory 
(\cpt), to relate it to  the \ktpp matrix element \cite{Bernard:wf}. Since lowest 
order $ SU(3) $ \cpt  is known to be accurate only at the $ 30 \% $ 
level, the step from \ktp to \ktpp induces large 
uncertainties. Unfortunately, the usage of a one loop \cpt  relation 
is not possible since this would imply the use of several low energy 
constants which are not available \cite{Kambor:1991bc}.

A crucial point for the calculation of the \ktpp 
matrix element is the inclusion of final state interaction (FSI), 
whose importance in the context of $ \varepsilon'/\varepsilon $ was pointed 
out by Pallante and Pich 
\cite{Pallante:1999qf,Pallante:2000hk}, following the ideas outlined in 
a paper of Truong \cite{Truong:1988zp}, who showed that the inclusion of FSI for 
\ktpp decays yields an enhancement of the $ I = 0 $ amplitude, pointing in 
the right direction concerning the 
$ \Delta I = 1/2 $ rule . These FSI are totally neglected if one 
relates the \ktp amplitude with the \ktpp amplitude with 
the help of the tree level \cpt relation.
In order to to write down a 
dispersion relation for the \ktpp amplitude, 
Pallante and Pich 
\cite{Pallante:1999qf}  
use an offshell kaon field, which can be defined in infinitely many ways, 
introducing also ambiguities in the final numerical result 
for the \ktpp amplitude. 
A detailed discussion of this point 
may be found in ref. \cite{Buchler:2001np}.

\section{Dispersive treatment with momentum carrying Hamiltonian}

One can avoid the problems related to the use of an offshell kaon field
by allowing the weak Hamiltonian to carry momentum; a procedure 
which has been suggested in ref. \cite{Buchler:2001nm}. We will sketch how 
the method works: Define the amplitude 
\be 
_{I=0} \langle \pi(p_1) \pi(p_2) | {\cal H}^{1/2}_W(0)|K(q_1) \rangle =: T^+(s,t,u) \co
\label{eq:Adef}
\ee 
with the Mandelstam variables 
$s=(p_1\!+\!p_2)^2, \;t=(q_1\!-\!p_1)^2, \; u=(q_1\!-\!p_2)^2$,
related by $s\!+\!t\!+\!u=2M_\pi^2\!+\!M_K^2\!+\!q_2^2$, where $q_2$ is the
momentum carried by the weak Hamiltonian. The physical decay amplitude is
obtained by setting $q_2^\mu=0$ ($s=M_K^2$, $t=u=M_\pi^2$).
To describe a function of three variables dispersively would be very complicated. 
The problem can be simplified considerably if we neglect the contribution of the 
imaginary parts of D waves and higher. In this 
approximation, the amplitude decomposes into several functions depending each only 
on one of the three Mandelstam variables: 

\bea
T^+(s,t,u)&=&M_0(s)+\left\{ {1 \over 3}\left[N_0(t)+2 R_0(t)\right]
\right. \no
&+& \left. {1 \over 2}\left[ \left(\!s-u-{M_\pi^2 \Delta \over t} \!\right)
  N_1(t)\right] \right\} \no
&+& \Big\{ (t \leftrightarrow u) \Big\} \co
\label{eq:T+dec}
\eea
where $\Delta = M_K^2-M_\pi^2$. $ M_0(s) $ corresponds to $ I=0 $ $ S $-wave in 
the $ s $ channel, whereas in the $ t $ channel  $ N_{0} $ and $ N_{1} $ denote 
the $ I= 1/2 $ $ S $ and $ P $ wave and $ R_{0} $ the $ I =3/2 $ $ S $ wave. 

The dispersion relation of the full amplitude is converted into a set of coupled 
dispersion relations of functions of a single variable, which can be solved 
numerically. For instance, for $M_0(s)$, giving the major 
contribution in the final result, we define 
the right-hand cut: 
\begin{eqnarray*}
{\rm disc} M_0(s) &=&\sin \delta_0^0(s) e^{-i\delta_0^0} \left[
  M_0(s)+\hat M_0(s) \right]   \co
\end{eqnarray*}
and get the dispersive representation: 

\begin{eqnarray*}
M_0(s) &=& \Omega_0^0(s,s_0)\left\{\rule{0em}{1.7em}a+b(s\!-\!s_0)\right. \\
&& \left.+{(s\!-\!s_0)^2 \over
    \pi} \int_{4M_\pi^2}^{\Lambda_1^2} {\sin \delta_0^0(s') \hat M_0(s') ds'
    \over |\Omega_0^0(s',s_0)| (s'\!-\!s)(s'\!-\!s_0)^2} \right\} \, , 
\end{eqnarray*}
with two subtraction constants (SC) $ a $ and $ b $. 
\newpage
\noindent The Omn$\grave{\mbox{e}}$s function $ \Omega_0^0(s,s_0) $ is defined to be: 

\begin{eqnarray*}
\Omega^0_0(s,s_0) = \exp\left\{{(s-s_0) \over \pi}
  \int_{4M_\pi^2}^{\tilde\Lambda_1^2} ds' 
  {\delta^0_0(s') \over (s'-s_0) (s'-s)} \right\} \fs
\end{eqnarray*}
$ \hat{M}_{0}(s) $ is an angular average of $ M_{0}(s) $. Likewise, one can define 
the same quantities for the remaining functions $ N_{0,1} $ and $ R_{0} $,  
but no more new SC's have to be implemented 
\cite{Buchler:2001nm}.
 
The number of SC's  which have to be introduced is crucial. In order to solve the 
dispersion relation and calculate the \ktpp amplitude, we have to provide two SC's  
as input. One of those can be linked to the \ktp amplitude: a soft pion theorem 
relates the amplitude at the soft pion point 
($ s = u = M_{\pi}^{2}, t = M_{K}^{2} $) to the \ktp amplitude up to order 
$ M_{\pi}^{2}$ corrections: 

\be
-{{\cal A}(K \to \pi)\over 2 F_\pi}= a + 
{ \bar N \over 3} +{\cal O}(M_\pi^2) \co 
\label{eq:soft pion point}
\ee 
where $\bar N=N_0(M_K^2)+2 R_0(M_K^2)$.
Notice that although the process involves a Kaon, the symmetry argument 
leading to the above relation is based on $ SU(2) $, and suffers therefore 
only from $ \mathcal{O}(M_{\pi}^{2}) $ corrections. With the help of 
this relation, the first subtraction constant can in principle be provided 
by a lattice calculation. The problem is the second subtraction constant b;
it is related to the derivative in $ s $ of the amplitude $ T^{+} $ at the 
soft pion point:  
\bes
b & = & {\partial \over \partial s} T^+(s,\Sigma-s,M_\pi^2)_{|s=M_\pi^2} + \ldots
\ees
In  ref. \cite{Buchler:2001nm} a Ward identity which relates this 
derivative to a \ktp matrix element is derived. The calculation of this matrix 
element would provide $ b $. 

Another option to get $ b $ is to calculate the amplitude (\ref{eq:Adef}) for 
the following unphysical kinematical values: 

\be
s=4 M_\pi^2,~~~ t=u={M_K^2 \over 2} - M_\pi^2 \co
\ee
where the two pions are produced at rest \cite{Dawson:1997ic}. This special kinematic configuration 
does not get into conflict with the no-go theorem of Maiani and Testa 
\cite{Maiani:ca}.

In the absence of a value for b, one can illustrate 
the numerical results by fixing it at a value and then varying it within a fairly 
wide range. For its central value one can use lowest order \cpt: 

\be b= {3 a\over M_K^2-M_\pi^2} \left(1+ X + {\cal
O}(M_K^4) \right) \fs
\label{eq:ab_CHPT}
\ee 
The size of the correction $ X $ is at the moment unknown, but nothing protects it 
of beeing of the order of $ M_{k}^{2} $.

\begin{figure}[t] 
\leavevmode \begin{center}
\includegraphics[width=7cm]{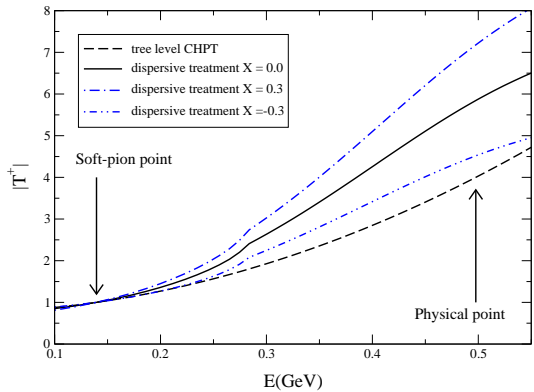}
\caption{\label{fig:kpp}The function $|T^+(s,t,u)|$ plotted {\em vs.}
  $E=\sqrt{s}$  along the line of constant $u=M_\pi^2$: the result of our
  numerical study for different values of $X$ are compared to tree level
  CHPT.}  
\end{center}
\end{figure}

The numerical analysis is shown in Fig.\ref{fig:kpp}, where 
$ T^{+}(s,M_{K}^{2}-M) $ as a function of the incoming Kaon momentum squared, $ s $, 
is plotted. 
For the uncertainty $ X $ coming from next to leading order \cpt, 
a rather wide range $ X = \pm 30\%$ has been chosen. 
Comparison with the lowest order \cpt formula, also plotted in Fig.\ref{fig:kpp}, 
shows that large corrections have to be expected due to the Omn$\grave{\mbox{e}}$s
factor, if we neglect next to leading order \cpt effects (X=0). If we vary $ X $
in the above given range, we see that we enhance the effect for positive $ X $
and decrease it for negative $ X $. In the case $ X = -0.3$ the 
Omn$\grave{\mbox{e}}$s-corrected curve differs only little from the 
lowest order \cpt one. 

\newpage
This analysis shows again that in order to get a reliable number for the \ktpp 
matrix element in this framework, it is crucial to obtain exact values for the 
two subtraction constants. The accuracy of the final result is essentially limited
by these, since the other potential source for uncertainties, 
the phase shifts needed as input for the Omn$\grave{\mbox{e}}$s functions, 
are known to a rather high precision.   

\section*{Acknowledgments}
It is a pleasure to thank G.~Colangelo, J.~Kambor and F.~Orellana for a pleasant 
collaboration on the subject discussed here.


\section*{References}
\begin{thebibliography}{99}

\bibitem{Lellouch:2000pv}
L.~Lellouch and M.~Luscher,
Commun.\ Math.\ Phys.\  {\bf 219}, 31 (2001)
[arXiv:hep-lat/0003023].

\bibitem{Maiani:ca}
L.~Maiani and M.~Testa,
Phys.\ Lett.\ B {\bf 245}, 585 (1990).

\bibitem{Bernard:wf}
C.~W.~Bernard, T.~Draper, A.~Soni, H.~D.~Politzer and M.~B.~Wise, 
Phys.\ Rev.\ D {\bf 32}, 2343 (1985).

\bibitem{Kambor:1991bc}
J.~Kambor,
TUM-T31-27-91
{\it Presented at Workshop on Effective Field Theories, Dobogoko, Hungary, Aug 22-26, 1991}.

\bibitem{Pallante:1999qf} 
E.~Pallante and A.~Pich,
Phys.\ Rev.\ Lett.\  {\bf 84}, 2568 (2000) 
[arXiv:hep-ph/9911233]. 

\bibitem{Pallante:2000hk}
E.~Pallante and A.~Pich,
Nucl.\ Phys.\ B {\bf 592}, 294 (2001)
[arXiv:hep-ph/0007208].

\bibitem{Truong:1988zp}
T.~N.~Truong,
Phys.\ Rev.\ Lett.\  {\bf 61}, 2526 (1988).

\bibitem{Buchler:2001np}
M.~Buchler, G.~Colangelo, J.~Kambor and F.~Orellana,
Phys.\ Lett.\ B {\bf 521}, 29 (2001)
[arXiv:hep-ph/0102289].

\bibitem{Buchler:2001nm}
M.~Buchler, G.~Colangelo, J.~Kambor and F.~Orellana,
Phys.\ Lett.\ B {\bf 521}, 22 (2001)
[arXiv:hep-ph/0102287].

\bibitem{Dawson:1997ic}
C.~Dawson, G.~Martinelli, G.~C.~Rossi, C.~T.~Sachrajda, S.~R.~Sharpe, M.~Talevi and M.~Testa,
Nucl.\ Phys.\ B {\bf 514}, 313 (1998)
[arXiv:hep-lat/9707009].
 
\end{thebibliography}
\end{document}